\documentclass[final,11pt]{article}

\usepackage{natbib}
\usepackage[margin=1.2in]{geometry}

\usepackage{amsmath}
\usepackage{amsfonts}
\usepackage{algorithm}
\usepackage{algorithmic}
\usepackage{color}
\usepackage{graphicx}
\usepackage{bmpsize}
\usepackage{times}
\usepackage[]{authblk}

\begin{document}

\global\long\def\J{J}
\global\long\def\x{\text{x}}
\global\long\def\s{\text{s}}
\global\long\def\t{\text{t}}
\global\long\def\u{\text{u}}
\global\long\def\r{\text{r}}
\global\long\def\p{\text{p}}
\global\long\def\q{\text{q}}
\global\long\def\w{\text{w}}
\global\long\def\KL{\text{KL}}
\global\long\def\T{\text{T}}
\global\long\def\M{\text{M}}

\newcommand{\traj}{\ensuremath{_{\t+1:\T}}}
\newcommand{\trajnull}{\ensuremath{_{1:\T}}}

\newcommand{\vnote}[1]{{\color{blue}[VG: #1]}} 
\newcommand{\dnote}[1]{{\color{magenta}[DT: #1]}} 

\newcommand{\avgp}{\ensuremath{\p}}
\newcommand{\avgustar}{\ensuremath{\u_\KL^{\ast}}}


\title{Action selection in growing state spaces: Control of Network Structure Growth}

\author[1]{\Large Dominik Thalmeier}
\author[2]{\Large Vicen\c{c} G\'omez }
\author[1]{\Large Hilbert J. Kappen }

\affil[1]{\large Donders Institute for Brain, Cognition and Behaviour\authorcr
       Radboud University Nijmegen, the Netherlands
       }
\affil[2]{\large Department of Information and Communication Technologies\authorcr
       Universitat Pompeu Fabra. Barcelona, Spain
       }
\date{}




\maketitle

\begin{abstract}
{\fontfamily{times}\selectfont
The dynamical processes taking place on a network depend on its topology.
Influencing the growth process of a network therefore has important implications on such dynamical processes.
We formulate the problem of influencing the growth of a network as a stochastic optimal control problem
in which a structural cost function penalizes undesired topologies.
We approximate this control problem with a restricted class of control problems that can be solved using probabilistic inference methods.
To deal with the increasing problem dimensionality, we introduce an adaptive importance sampling method
for approximating the optimal control.
We illustrate this methodology in the context of formation of information cascades,
considering the task of influencing the structure of a growing conversation thread, as in Internet forums.
Using a realistic model of growing trees, we show that our approach 
can yield conversation threads with better structural properties than the ones observed without control.}
\end{abstract}

%

%
%
%

\section{Introduction}
{\fontfamily{times}\selectfont
Many complex systems can be described as dynamic processes which are characterized by the topology of an underlying network. Examples of such systems are human interaction networks, where the links may represent transmitting opinions~\cite{olfati2007consensus,dai2011optimal,Centola17022015}, habits~\cite{Centola03092010,farajtabar2014shaping}, money~\cite{gai2010contagion,amini2013resilience,giudici2016graphical} or viruses~\cite{PhysRevLett.86.3200,PhysRevLett.89.108701}. Being able to control, or just influence in some way, the dynamics of such complex networks may lead to important progress, for example, avoiding financial crises, preventing epidemic outbreaks or maximizing information spread in marketing campaigns.

The control of the dynamics on networks is a very challenging problem that has attracted significant interest recently~\cite{liu2011controllability,Cornelius2013,gao2014target,Yan2015}.
Existing approaches typically consider network controllability as the controllability of the dynamical system induced by
the underlying network structure.
While it is agreed that network controllability critically depends on the network structure,
the problem of how to control the network structure itself while it is evolving remains open.

The network structure is determined by the dynamics of addition and deletion of nodes and links over time.
In this paper, we address the problem of influencing this dynamics in the framework of stochastic optimal control. 
The standard way to address these problems is through the Bellman equation and dynamic programming.
Dynamic programming is only feasible in small problems and requires approximations when the 
state and action spaces are large. In the setting of network growth, this problem is more severe, since 
the state space increases (super-)exponentially with the number of nodes.

In order to deal with this curse of dimensionality, we propose to approximate the network growth control problem
by a special class of stochastic optimal control problems, known as Kullback-Leibler~(KL) control or
Linearly-Solvable Markov Decision Problems (LMDPs) \cite{TodorovPNAS2009,KappenML2012}.
For this class of problems, one can use efficient adaptive importance sampling methods 
that scale well in high dimensions.
The optimal solution for the KL-control problem tends to be sparse, so that only a few
\emph{next} states become relevant, effectively reducing the branching factor of the original problem.
The obtained solution of the KL-control problem is then used to compute the optimal action in the
original problem that does not belong to the KL-control class.


In the next section we present our proposed general methodology.
We then apply it to a realistic problem: influencing the growth process of cascades in online forums,
in order to maximize structural network measures that are connected to the quality of an online conversation thread.
We conclude the paper with a discussion.
}


\section{Optimal Network Growth as a Control Problem}

\label{sec:framework} We now formulate the network
growth control problem as a stochastic optimal control problem. Let 
$\x_{\t}\in\mathcal{X}$, with $\mathcal{X}$ being the set of all possible network structures, denote the growing structure (state) of the network at time $\t$
 and let $P(\x'|\x,\u)$ describe the network dynamics, where the control variable $\u\in\mathcal{U}$
denotes possible actions we can perform in order to manipulate the network.
Let us label the default action, which means not interacting with the system, with $\u=0$.
We denote the corresponding dynamics without control as the uncontrolled process $\p(\x'|\x):= P(\x'|\x,\u=0)$.

At each time-step $\t$, we incur an arbitrary cost function on the network state $\r(\x,\t)$ which is assigned when the state is reached.
The state cost $\r(\x,\t)$ penalizes network structures that
are not convenient in the particular context under consideration.
For example, if one wants to favour networks with large average clustering coefficient $\text{\text{C}}(\x)$,
then $\r(\x,\t)=-\text{C}(\x)$. Alternatively, one can consider
more complex functions, such as the structural virality or Wiener
index~\cite{WienerMohar}, as proposed recently \cite{goel2015structural},
to maximize the influence in a social network. In general, any measure
that can be (efficiently) computed from $\x$ fits the presented framework.

Our objective is to find the control function $\u(\x,\t):\mathcal{X}\times\mathbb{N}\mapsto\mathcal{U}$
 which minimizes the total cost over a time horizon $\T$, starting at state $\x$ at initial time $\t=0$
\begin{equation}
\mathcal{C}\left(\x,\t=0,\u(\cdot)\right)=\r\left(\x,0\right)+\left\langle \sum_{\t'=\t+1}^{\T}\r\left(\x_{\t'},\t'\right) \right\rangle _{P(\x\trajnull|\x,\u(\cdot),\t=0)},\label{eq:totalCost}
\end{equation}
where the expectation is taken with respect to the probability $P(\x\trajnull|\x,\u(\cdot),\t=0)$
over paths $\x\trajnull$ in the state space, given state $\x$ at time $\t=0$ using the control-function $\u(\cdot)$.
The probability of a path is given by $P(\x\traj|\x,\u(\cdot),\t=0)=\prod_{\s=\t}^{\T-1}P(\x_{\s+1}|\x_{\s},\u(\x_{\s},\s),\s)$.

Computing the optimal control can be done by dynamic programming~\cite{bertsekas1995dynamic}.
We introduce the optimal cost-to-go
\begin{equation}
\ensuremath{\J}(\x,\t)=\min_{\u(\cdot)}\mathcal{C}\left(\x,\t,\u(\cdot)\right),
\end{equation}
which is an expectation of
the cumulative cost starting at state $\x$ and time $\t$ and acting optimally thereafter. This can be computed using the Bellman equation
\begin{equation}
\ensuremath{\J}(\x,\t)=\ensuremath{\min}_{\u}\left(\r(\x,\t)+\left\langle \J(\x',\t+1)\right\rangle _{P(\x'|\x,\u,\t)}\right).\label{eq:bellman1}
\end{equation}
From $\J(\x,\t)$, the optimal control is obtained by a greedy local optimization:
\begin{equation}
\u^{\ast}(\x,\t)=\text{argmin}_{\u}\left(\r(\x,\t)+\left\langle \J(\x',\t+1)\right\rangle _{P(\x'|\x,\u,\t)}\right).\label{eq:ustar}
\end{equation}
In general, the solution to equation \eqref{eq:bellman1} can be computed 
recursively using dynamic programming \cite{bertsekas1995dynamic}
for all possible states. This is however infeasible for controlling
network growth, as the computation is of polynomial order in the number of states and the state
space of networks increases super-exponentially on the number of nodes. E.g. for directed unweighed networks, there
are $2^{\text{N}^{2}}$ possible networks with $\text{N}$ labelled nodes.


\section{Approximating the network growth problem by a Kullback-Leibler control problem}
\label{sec:KL-control}
In this section we present our main approach, which first computes the optimal cost-to-go
on a relaxed problem and then uses it as a proxy for the original optimal cost-to-go.
In the next subsection, we introduce the class of KL-control problems that we use as a relaxation.
We then illustrate KL-control using a tractable example of tree growth.
In subsection~\ref{sec:sampling}, we explain 
how can we approximate the KL-control solution using the cross-entropy method.
Finally, in subsection~\ref{sec:limited} we show how can we use that result to compute the action
selection in the original problem.
\subsection{Kullback-Leibler control}
In order to efficiently compute the optimal cost-to-go, we make the assumption that 
our controls directly specify the transition probabilities between two subsequent network structures, e.g.~$P(\x'|\x,\u(\t))\approx\u(\x'|\x,\t)$.
Further, we define the natural growth process of the network (the uncontrolled dynamics) as a Markov chain
with transition probabilities $\p(\x'|\x)$.
Because our influence on the network dynamics is limited, we add a regularization term to the total cost defined in equation \eqref{eq:totalCost} that penalizes deviations from $\p(\x'|\x)$.
The approximated control cost becomes
\begin{align}
\mathcal{C}_\KL^\lambda\left(\x,\t,\u(\cdot)\right)=&\lambda\KL\left[\u\left(\x\traj|\x,\t\right)\parallel\p\left(\x\traj|\x,\t\right)\right]
+\r\left(\x,\t\right)+\left\langle \sum_{\t'=\t+1}^{\T}\r\left(\x_{\t'},\t'\right) \right\rangle _{\u\left(\x\traj|\x,\t\right)},\label{eq:totalCostKL}
\end{align}
with the $\KL$-divergence
\[
\text{KL}\left[\u\left(\x\traj|\x,\t\right)\parallel\p\left(\x\traj|\x,\t\right)\right]=\left\langle \log\frac{\u\left(\x\traj|\x,\t\right)}{\p\left(\x\traj|\x,\t\right)}\right\rangle _{\u\left(\x\traj|\x,\t\right)},
\]
which measures the closeness of the two path distributions, $\p\left(\x\traj|\x,\t\right)$ and $\u\left(\x\traj|\x,\t\right)$. The parameter $\lambda$ thereby regulates the strength of this penalization.

With this assumption, the control problem consisting in minimizing $\mathcal{C}_\KL^\lambda$ w.r.t. the control $\u(\x'|\x,\t)$ belongs to the KL-control class and has a closed form solution~\cite{TodorovPNAS2009,KappenML2012}.
The probability distribution of an optimal path $\u^{\ast}_\KL\left(\x\traj|\x,\t\right)$ that minimizes equation \eqref{eq:totalCostKL} is
\begin{equation}
\u^{\ast}_\KL\left(\x\traj|\x,\t\right)=\frac{\p\left(\x\traj|\x,\t\right)}{\left\langle \phi(\x\traj)\right\rangle _{\p(\x\traj|\x,\t)}}\phi(\x\traj),\label{eq:optpath}
\end{equation}
with
\begin{equation}
\phi(\x\traj):=\exp\left(-\lambda^{-1}\sum_{\t'=\t+1}^{\text{T}}\r(\x_{\t'},\t')\right).\label{eq:phi}
\end{equation}

Plugging this into equation~\eqref{eq:totalCostKL} and minimizing gives the optimal cost-to-go
\begin{equation}
\J^\lambda_\KL(\x,\t)=\r(\x,\t)-\lambda\log \left\langle \phi(\x\traj)\right\rangle _{\p(\x\traj|\x,\t)},\label{eq:J_solution}
\end{equation}
which can be numerically approximated using paths sampled from the uncontrolled dynamics $\p(\x\traj|\x,\t)$. 

The optimal control corresponding to equation~\eqref{eq:ustar} corresponds to a
state transition probability distribution that is obtained by marginalization in equation~\eqref{eq:optpath}.
It is expressed in terms of the uncontrolled transition probability $\p(\x'|\x)$ and the (exponentiated) optimal cost-to-go:
\begin{equation}
\u^{\ast}_\KL(\x'|\x,\t)=
\sum_{\x_{\t+2:\T}}\u^{\ast}_\KL\left(\x_{\t+1}=\x',\x_{\t+2:\T}|\x,\t\right)\\
\propto
\p(\x'|\x)\exp\left(-\frac{\J^\lambda_\KL(\x',\t+1)}{\lambda}\right).\label{eq:optcontrol}
\end{equation}
This resembles a Boltzmann distribution with temperature $\lambda$ where the optimal cost-to-go takes the role of an energy.
The effect of the temperature becomes clear: for high values of $\lambda$, $\u^{\ast}_\KL(\x'|\x,\t)$ deviates only a little from the uncontrolled dynamics $\p(\x'|\x)$, thus the optimal control has a weak influence on the system.
In contrast, for low values of $\lambda$, the exponential in equation \eqref{eq:optcontrol} becomes very pronounced for the state(s) $\x'$ with the smallest cost-to-go $\J^\lambda_\KL(\x',\t+1)$, suppressing the transition probabilities to suboptimal states $\x'$. Thus the control has a very strong effect on the process. In the limit of $\lambda$ going to zero, the controlled process becomes deterministic, if $\J^\lambda_\KL(\x',\t+1)$ is not degenerate (meaning there is a unique state $\x'$ which minimizes the optimal cost-to-go). In this case the control is so strong that it overpowers the noise completely.

We thus approximate our original (possibly difficult) control problem as a $\KL$-control problem,
parametrized by the temperature $\lambda$.
The approximated optimal cost-to-go $\J(\x',\t+1)$ of equation \eqref{eq:ustar} 
is replaced by the corresponding optimal cost-to-go of the KL-control problem
$\J^\lambda_\KL(\x',\t+1)$ of equation~\eqref{eq:optcontrol} and used to compute the action selection in
the original problem.


\subsection{A Tractable Example}
\label{sec:toy}

We now present a tractable example amenable for exact optimal control computation.
This example already belongs to the $\KL$-control class, so no approximation is made.
The purpose of this analysis it to show how different values of the temperature $\lambda$ 
may lead to qualitatively different optimal solutions and other interesting phenomena.

Let's consider a tree that grows at discrete time-steps, starting with the root node
at time $\t=0$. We represent the tree at time $\t$ as a vector $\x_{\t}=(x_0,x_1,...,x_\t)$,
where $x_\t$ indicates the label of the parent of the node attached at time $\t$.
At every time-step, either the tree remains the same or a new node is attached to it.
The root node has label $1$ and the label $0$ is specially used to indicate that no node was added at a given time-step
(it is also the label of the parent of the root node).
The nodes are labelled in increasing order as they arrive to the tree, so that
at time-step $\t$, for a tree with $k$ nodes, $k\leq t, x_{\t}=0,1,\hdots,k$ corresponds to the parent of node $k+1$ if a node is added or zero otherwise.
Thus, the parent vector at time $\t=1$ is always $\x_1=(0,1)$.

Our example is a finite horizon task of $\T=10$ time-steps and end-cost only. The end-cost implements two control objectives: it prefers trees of large Wiener index while penalising trees with many nodes (more than five, in this case).
The Wiener index is the sum of the lengths of the shortest paths between all nodes in a graph. It is maximal for a chain and minimal for a star.

The uncontrolled process is biased to the root: new nodes choose to link the root with probability $3/5$ and uniformly otherwise.
More precisely
\begin{align}
\p(x_{t+1}=j|\x_{\t}) &= \begin{cases} \frac{3}{5}   &\text{for $j=1$}\\
								\frac{2}{5\|\x_{\t}\|_0} &\text{for $(j=0) $ or $ j \in \{2,\hdots,\|\x_{\t}\|_0 \}$}
								\end{cases}\label{eq:toy1}\\
\r(\x_{\t},\t) &= \begin{cases} -\mathtt{Wiener}(\x_{\t})\delta_{\t,\T} &\text{if $\|\x_{\t}\|_0<5$}\\
							\delta_{\t,\T} & \text{otherwise}
							\end{cases}\label{eq:toy2}
\end{align}
where $\|\x\|_0$ denotes the number of non-zero elements in $\x$ and $\texttt{Wiener}$ the (normalized) Wiener index.

In this setting, the uncontrolled process $\p$ tends to grow trees with more than five nodes with many of them attached to the root node, i.e. with low Wiener index. We want to influence this dynamics so that the target configuration, a chain of five nodes (maximal Wiener index) is more likely to be obtained.
\begin{figure*}[!t]
\centering
\includegraphics[width=.8\textwidth]{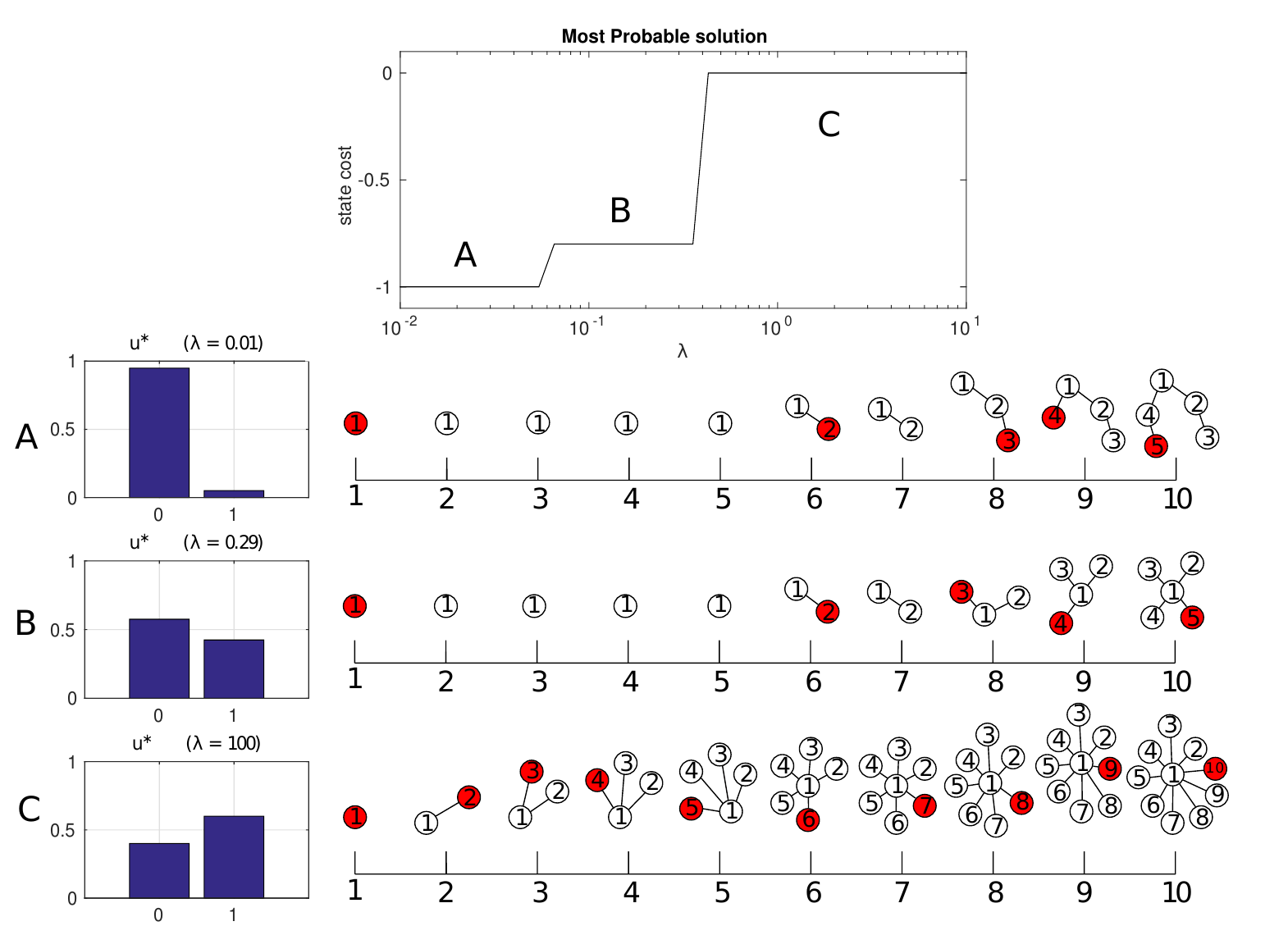}
\caption{\label{fig:example} 
Example of optimal control of tree growth. \textbf{(Top)}: the state cost of the most probable solutions as a function of the temperature $\lambda$.
In region A, the optimal strategy waits until the last time-steps and then grows a tree with maximal Wiener index.
In region B, it builds a star of five nodes.
Finally, in region C, it follows the uncontrolled dynamics and builds a star of ten nodes.
  \textbf{(Left)}: for each region, the optimal probabilities $\u^*(x_{\t+1}|\x_{\t})$ at $\t=1$ for the two actions which are initially available: no node addition (0) and adding a node to the root (1).
In regions A, B the optimal control favours not adding a new node initially.
The sequences on the \textbf{right} show how the tree grows. When a new node is added to the tree, it is coloured in red.
}
\end{figure*}

Figure~\ref{fig:example} (top) shows the state cost $\r$ of the final tree that results from choosing the most probable control (MAP solution) as a function of the temperature $\lambda$. 
The exact solution is calculated using dynamic programming~\cite{KappenML2012}. We can differentiate three types of solutions, denoted as A, B and C in the figure.

For low temperatures (region A) the control aims to fulfil both control objectives: to find a small network with maximal Wiener index. The optimal strategy does not add nodes initially and then builds a tree of maximal Wiener index (see inset of initial controls in left column of the figure).
This type of control (to wait while the target is far in the future) is reminiscent of the delayed choice mechanism described previously~\cite{kappen_prl05}.
This initial waiting period makes sense because if the chain of length $5$ would be grown immediately, then at time $6$ the size of $5$ is already reached. If now an additional node attaches, then the final cost would be zero. However if one first waits and then grows the chain, an accidental node insertion before time $6$ would not be so disastrous (actually it may help), as one can then just wait until time $7$ to start growing the rest of the tree. So delaying the decision when to start growing the tree helps compensating accidental events.

For intermediate temperatures (region B), the initial control becomes less extreme, as we observe if we compare the left plots between regions A and B. For $\lambda\approx 0.07$, the solution that builds the tree with maximal Wiener index is no longer optimal, since it deviates too much from the uncontrolled dynamics. In region B, the control aims to build a network of five nodes or less,  but no longer aims to maximize the Wiener index. The control is characterized by an initial waiting period and the subsequent growth of a tree of five nodes, which are in this case all attached to the root node.  

Finally, for high temperatures (region C, $\lambda> 0.4$), the control essentially ignores the cost $\r$ and the optimal strategy is to add one node to the root at every time-step, following the uncontrolled process.

From these results we conclude that KL-control as a mechanism for controlling network growth
can capture complex phenomena such as transitions between qualitatively different optimal solutions and delayed choice effects.

\subsection{Sampling from the $\KL$-optimally controlled dynamics}
\label{sec:sampling}
In this subsection, we explain how we can sample from the optimally controlled dynamics and thereby obtain an estimate of the optimal cost-to-go $\J^\lambda_\KL(\x,\t)$ of equation~\eqref{eq:J_solution}.

The probability of an optimally controlled path, equation~\eqref{eq:optpath}, corresponds to the product of the uncontrolled dynamics by the exponentiated state costs. Hence a naive way to obtain samples from the optimal dynamics, would consist in sampling paths from the uncontrolled dynamics $\p(\x'|\x)$ and weight them by their exponentiated state costs.
Using these samples we can then compute expectations from the optimally controlled dynamics. 
We use that for any function $f(\x\traj)$ we have:
\[
\left\langle f(\x\traj) \right\rangle _{\u_\KL^{\ast}(\x\traj|\x,\t)} = \left\langle f(\x\traj)\frac{\phi(\x\traj)}{\left\langle \phi(\x\traj)\right\rangle _{\p(\x\traj|\x,\t)}} \right\rangle _{\p(\x\traj|\x,\t)}.
\]

More precisely, provided a learned model or a simulator of the uncontrolled dynamics $\p(\x'|\x)$, we generate $\M$ sample paths $\x\traj^{(i)}, i=1,\hdots,\M$ from $\p(\x'|\x)$ and compute the weights $\frac{\phi(\x\traj^{(i)})}{\hat{\phi}}$.
The denominator thereby gives with equation \eqref{eq:J_solution} an estimate of the optimal cost-to-go as
\begin{align*} 
\left\langle \phi(\x\traj)\right\rangle _{\p(\x\traj|\x,\t)}& \approx\hat{\phi}:=\frac{1}{\M}\sum_{i=1}^{\text{M}}\phi\left(\x^{(i)}\traj\right).
\end{align*}
This method can be combined with resampling techniques~\cite{douc2005comparison,hol2006resampling} to obtain unweighted samples $\x\traj^{\text{opt},(i)}$ from the optimal dynamics (for the numerical methods in this article, we have used structural resampling~\cite{douc2005comparison,hol2006resampling}).

Using such a naive sampling method, however, can be inefficient, specially for low temperatures.
While for high temperatures $\lambda$ basically all weights $\frac{\phi(\x\traj^{(i)})}{\hat{\phi}}$ are more or less equal,
for low temperatures only a few samples with very large weights contribute to the approximation, 
resulting in very poor estimates.

This is a standard problem in Monte Carlo sampling and can be addressed using the Cross-Entropy (CE)
method~\cite{de2005tutorial,kappen2015adaptive}, which is an adaptive importance sampling 
algorithm that incrementally updates a baseline sampling policy or sequence of controls.
Here we propose to use the CE method in the discrete formulation and use 
a parametrized Markov process $\widetilde{\u}_{\omega}(\x'|\x,\t)$, with parameters $\omega$, to approximate $\u_\KL^\ast$.
The CE method in our setting alternates the following steps:
\begin{enumerate}
\item
In the first step, the optimal control is estimated using $\text{M}$ sample paths drawn from a
parametrized proposal distribution $\widetilde{\u}_{\omega}(\x'|\x,\t)$.
\item
In the second step, the parameters $\omega$ are updated so that
the proposal distribution becomes closer to the optimal probability distribution.
\end{enumerate}

As a proposal distribution $\widetilde{\u}_{\omega}(\x'|\x,\t)$, we use
\begin{equation}
\widetilde{\u}_{\omega}(\x'|\x,\t)\propto\p(\x'|\x)\exp\left(-\frac{\widetilde{\J}_\KL(\x',\omega(\t))}{\lambda}\right),\label{eq:impsampler}
\end{equation}
which has the same functional form as the optimally controlled transition probabilities in equation \eqref{eq:optcontrol}.
The $\KL$-optimal cost-to-go is thereby approximated by a linear sum of time-dependent feature
vectors $\psi_{k}^{\t}(\x)$
\begin{equation}
\widetilde{\J}_\KL(\x,\omega(\t))=\sum_{k}\omega_{k}(\t)\psi_{k}^{\t}(\x). \label{eq:Jparam}
\end{equation}

The probability distribution of an optimally controlled path, equation~\eqref{eq:optpath}, can be written as
\begin{equation}
\u^{\ast}_\KL\left(\x^{(i)}\traj|\x,\t\right)\propto\widetilde{\u}_{\omega}(\x^{(i)}\traj|\x,\t)\frac{\p\left(\x\traj^{(i)}|\x,\t\right)}{\widetilde{\u}_{\omega}(\x^{(i)}\traj|\x,\t)}\exp\left(-\lambda^{-1}\sum_{\t'=\t+1}^{\text{T}}\r(\x^{(i)}_{\t'},\t')\right).\label{eq:optpath2}
\end{equation}

This shows that we can draw samples from the proposal distribution and reweight them with the combined weights
\[
\w^{(i)}=\frac{\p(\x^{(i)}\traj|\x,\t)}{\widetilde{\u}_{\omega}(\x^{(i)}\traj|\x,\t)}\phi\left(\x^{(i)}\traj\right).
\]

The parameters $\omega_{k}(\t)$ of the importance sampler are initialized with zeros, which 
makes the initial proposal distribution equivalent to the uncontrolled dynamics.
The procedure requires the gradients of $\widetilde{\u}_{\omega}(\x'|\x,\t)$ at each iteration.
We describe the details of the CE method in~\ref{sec:tachan}.

We measure the efficiency of an obtained proposal control using the effective sample size (EffSS),
which estimates how many effective samples can be drawn from the optimal distribution. 
Given $\M$ samples with weights $\w^{(i)}$, the EffSS is given by
\begin{equation}
\text{EffSS}=\frac{\frac{1}{\text{M}}\sum_{i=1}^{\text{M}}\left(\w^{(i)}\right)^2}{\left(\frac{1}{\text{M}}\sum_{i=1}^{\text{M}}\w^{(i)}\right)^{2}}. \label{eq:EffSS}
\end{equation}
If the weights $\w^{(i)}$ are all about the same value, the EffSS is high, indicating that many samples contribute to statistical estimates using the weighted samples. If all weights are equal, the EffSS is equal to the number of samples $\M$. Conversely if the weights $\w^{(i)}$ have a large spread, the EffSS is low, indicating that only few independent samples contribute to statistical estimates. In the extreme case, when one weight is much larger then all others, the EffSS approaches $1$.
\subsection{Action selection using the KL-approximation}
\label{sec:limited}
Once we have an estimate of the cost-to-go $\J^\lambda_\KL$, we need to select an action $\u\in\mathcal{U}$ in the original control problem, which is not of the $\KL$-control type.
We select the optimal action according to
\begin{equation}
\u^{\ast}(\x,\t)\approx\text{argmin}_{\u}\left(\text{r}(\x,\t)+\left\langle \J^\lambda_\KL(\x',\t+1)\right\rangle _{P(\x'|\x,\u,t)}\right),\label{eq:ustarproxy}
\end{equation}
which requires 
the computation of $\J^\lambda_\KL(\x_{\t+1},\t+1)$ for every 
reachable state $\x_{\t+1}$.
In growing networks, the number of possible next states (the branching factor)
increases quickly, and visiting all of them soon becomes infeasible.

In this subsection we highlight an important benefit of using the KL-approximation as a relaxation of the original problem: the optimally controlled process tends to discard many irrelevant states, specially for small values of $\lambda$. This means that $\u_\KL^{\ast}(\x'|\x,\t)$ is sparse on $\x'$ (only a few next states are relevant for the task),
since the cost $\J^\lambda_\KL(\x',t)$ is very large for the corresponding $\x'$ where $\u_\KL^{\ast}(\x'|\x,\t)\approx 0$. 

Let $\x\traj^{\text{opt}}$ denote a trajectory sampled from the optimally controlled process, as described in the previous section.
We compute $\u_\KL^{\ast}(\x'|\x,\t)$ using:
\begin{equation}
\hat{\u}_\KL^{\ast}(\x'|\x,\t)= \left\langle \delta_{\x^{\text{opt}}(\t+1),\x'} \right\rangle _{\u_\KL^{\ast}(\x\traj|\x,\t)},
\end{equation}
where $\x^{\text{opt}}(\t+1)$ is the first element of the trajectory and $\delta_{\x^{\text{opt}}(\t+1),\x'}$ is the Kronecker delta which is equal one if $\x^{\text{opt}}(\t+1)$  is equal to $\x'$, and zero otherwise.

We then compute the optimal cost using equation~\eqref{eq:optcontrol}:
\begin{equation}
\J^\lambda_\KL(\x',\t+1)\sim -\log\left(\frac{\hat{\u}_\KL^{\ast}(\x'|\x,\t)}{\p(\x'|\x,\t)}\right),\label{eq:jklhatu}
\end{equation}
where we dropped a term which does not depend on $\x'$ and therefore plays no role in the minimization of equation~\eqref{eq:ustarproxy}.
The KL-approximation can help reducing the branching factor
because it needs only a few samples to calculate $\J^\lambda_\KL(\x',\t+1)$ only for the $\x'$ where $\u_\KL^{\ast}(\x'|\x,\t) > 0$ and thus $\J^\lambda_\KL(\x',t)$ has a finite value.

As mentioned earlier,  $\u_\KL^{\ast}(\x'|\x,\t)$ tends to be more sparse for small values of $\lambda$, when the KL-control problem is less noisy. In~\ref{sec:analyzinglimited} we provide analytical details of the two extreme conditions, when $\lambda$ is zero or infinite, respectively.

\begin{figure}[t!]
\centering
\includegraphics[width=.8\columnwidth]{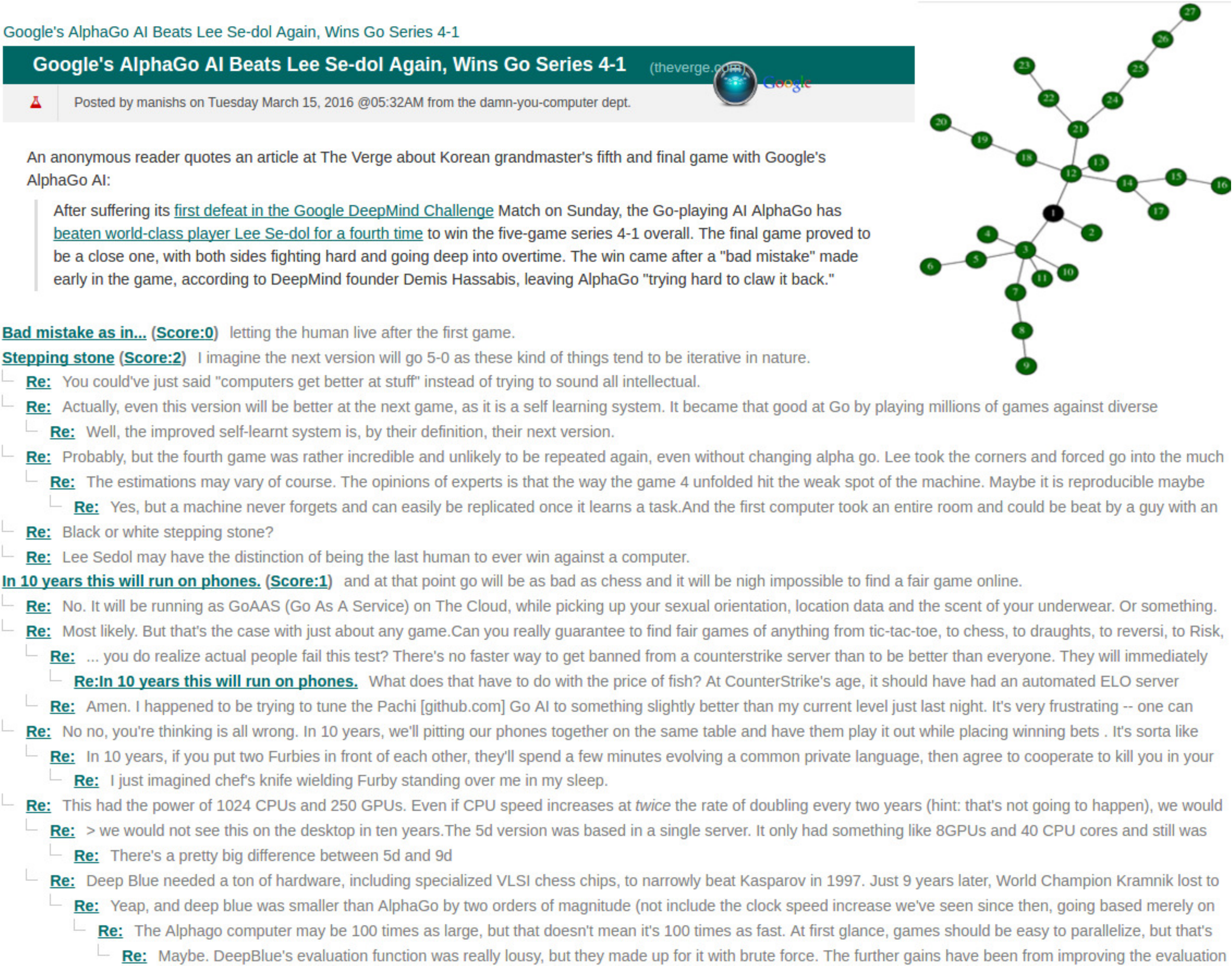}
\caption{\label{fig:task1} 
Task illustration: example of an Internet news forum.
News are posted periodically and users can write comments either 
to the original post or to other user's comments, forming a cascade of messages.
The figure shows an example of conversation thread taken from Slashdot about \emph{Google's AlphaGo}.
The control task is to influence the structure of the conversation thread (shown as a growing tree in the top-right).
}
\end{figure}

\section{Application to Conversation Threads}
\label{sec:threads}
We have described a framework for controlling growing graphs.
We now illustrate this framework in the context of growing information cascades.
In particular, we focus on the task of controlling the growth of online conversation threads.
These are information cascades that occur, for example, in online forums such as weblogs~\cite{leskovec2007patterns},
news aggregators~\cite{gomez2008statistical} or the synthesis of articles of Wikipedia~\cite{ICWSM112764}.
In conversation threads, after an initial post appears, different users react writing comments either
to the original post or to comments from other users.

Figure~\ref{fig:task1} shows an example of a conversation thread, taken from Slashdot (\texttt{www.slashdot.org}).
Users see a conversation thread using a similar hierarchical interface.

\begin{figure}[t!]
\centering
\includegraphics[width=.8\columnwidth]{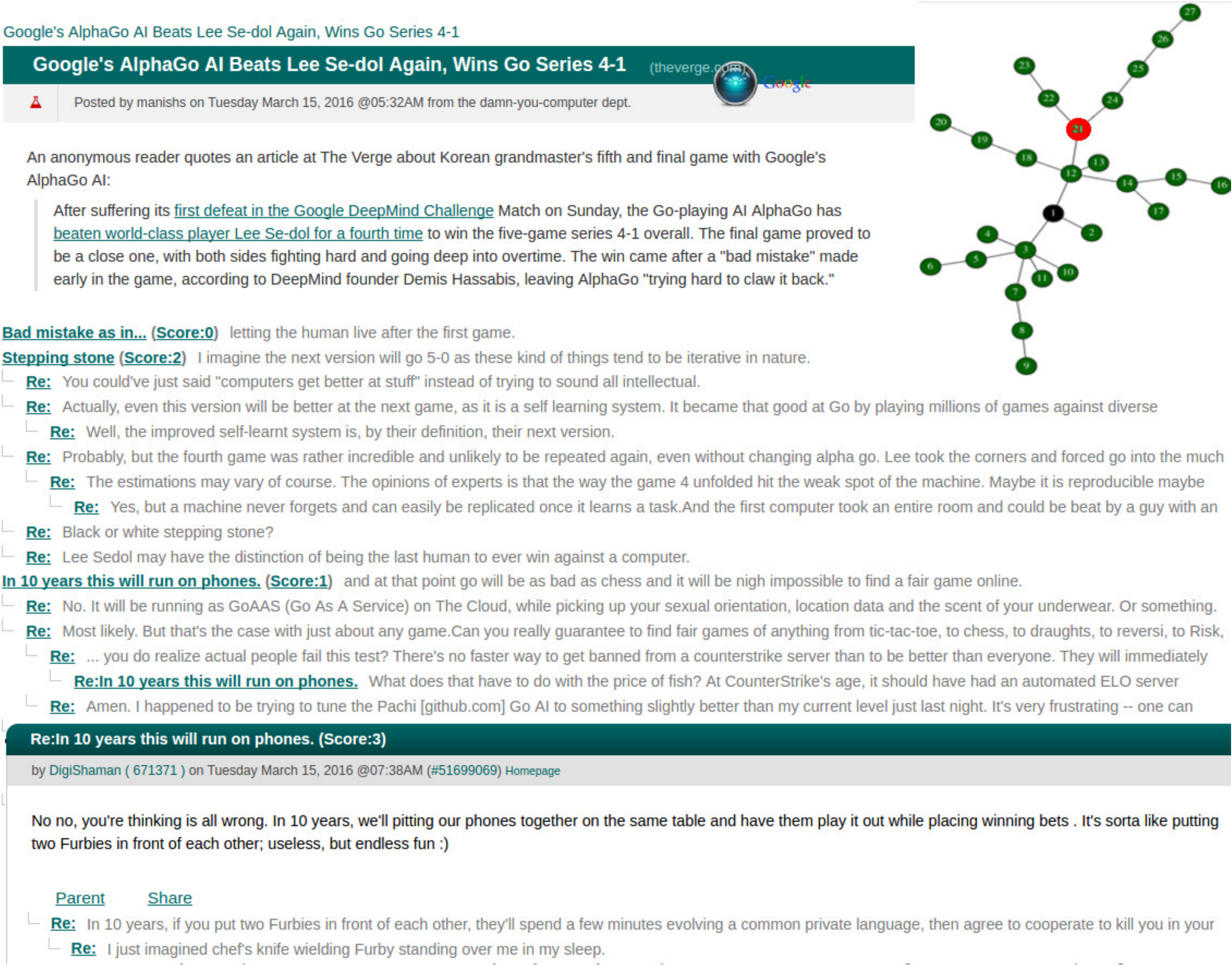}
\caption{\label{fig:task2} 
Our proposed control mechanism: in addition to the the threaded conversation,
we highlight a comment (red node in the growing tree), suggested to be replied by the user.
The choice of suggested comment, shown at the bottom of the page, is calculated using the method
described in section~\ref{sec:sampling}.
}
\end{figure}

The task we consider is to optimize the structure of the generated conversation thread while it grows.
The state is thus defined as a growing tree.
We assume an underlying (not observed) population of users that keep adding nodes to this tree.
Since we can not control directly what is the node that will receive the next comment, 
we propose the user interface as a control mechanism to influence indirectly the growth process.
This can be done in different ways, for example, manipulating the 
layout of the comments. In our case, the control signal will be to recommend a comment (by highlighting it) 
to which the next user can reply. Figure~\ref{fig:task2} illustrates such a mechanism.
The action selection strategy introduced in section~\ref{sec:limited} is used to select the comment to highlight.
Our goal is thus to modify the structure of a cascade in certain way while it evolves, by influencing its growth indirectly.
It is known that the structure of online threads is strongly related with the complexity of the underlying conversation~\cite{gomez2008statistical,gonzalez2010structure}. 

To fully define our control problem, we need to specify the structural cost function, the uncontrolled
dynamics, i.e. the equivalent of equations~\eqref{eq:toy1} and~\eqref{eq:toy2} for this task,
and a model of how an action (highlighting a node) changes the dynamics.
Globally, this application differs from the toy example of subsection~\ref{sec:toy} in some important ways:
\begin{enumerate}
\item The state-space is larger (threads typically receive more than $10$ comments).
\item We choose as state-cost function the Hirsch index (h-index), which makes the control task highly non-trivial.
\item The original problem is not a $\KL$-control problem. We use the action selection method described in section \ref{sec:limited} to control the growth of the conversation thread.
\end{enumerate}

\subsection{Structural Cost Function}
We propose to optimize the Hirsch index (h-index) as structural measure.
In our context, a cascade with h-index $h$ has $h$ comments each of which have received at least $h$ replies.
It is a sensible quantity to optimize, since it measures how distributed the comments of users on previous comments are.
A high h-index prevents two extreme cases that occur in a rather poor conversation: the case where a small number of posts
attract most of the replies, thus there is no interaction, 
and the case with deep chains, characteristic of a flame war of little interest for the community.
Both cases have a low h-index, while a high h-index spreads the conversation over multiple levels of the cascade.

The h-index is a function of the degree sequence of all nodes in the tree,
where the degree of a node is this case is the number of replies plus one, as there is also a link to the parent (replied comment or post). Therefore we use the degree histogram as features $\psi_k^\t(\x)$ for the parametrized form of the optimal cost-to-go, equation~\eqref{eq:Jparam}. That is,
feature $\psi_k^\t(\x)$ is the number of nodes with degree $k$ in the tree $\x$ at time-step $\t$.
We model the problem as a finite horizon task with end-cost.
Thus, the state cost is defined as $\r(\x,\t)=-\delta_{\t,\T}\cdot \text{h}(\x)$,
where $\text{h}(\x)$ is the h-index of the tree $\x$.

\subsection{Uncontrolled Dynamics for Online Conversation Threads}

As uncontrolled dynamics, we use a realistic model that determines the probability of a comment
to attract the replies of other users at any time, by means of an interplay between the following features:
\begin{itemize}
\item \emph{Popularity} $\alpha$: number of replies that a comment has already received.
\item \emph{Novelty} $\tau$: the elapsed time since the comment appeared in the thread.
\item \emph{Root node bias} $\beta$: characterizes the level of trendiness of the main post.
\end{itemize}
Such a model has proven to be successful in
capturing the structural properties and the temporal evolution of discussion threads present in
very diverse platforms~\cite{gomez13www}.
Notice that these features $\theta = (\alpha,\tau,\beta)$ should not be confused with the features
$\psi_k^\t(\x)$ used to encode the cost-to-go.

We represent the conversation thread as a vector of parents $\x_{\t}=(x_0,x_1,...,x_\t)$.
Given the current state of the thread $\x_{\t}$, the uncontrolled dynamics attaches a new node $\t+1$ to an existing
node $j$ with probability
\begin{align}
\p_\theta(x_{\t+1}=j|\x_{\t}) &= \frac{1}{Z_{\t+1}}\left(\text{deg}_{j,\t}\alpha+\delta_{j,1}\beta+\tau^{\t+1-j}\right) \label{eq:model}
\end{align}
with $Z_{\t+1}$ a normalization constant, $\text{deg}_{j,\t}$ the degree of node $j$ at time $\t$ and $\delta_{j,1}$ the Kronecker delta function,
so parameter $\beta$ is only nonzero for the root.

Given a dataset composed of S threads $\mathcal{D}:=\{\x^{(1)},\hdots,\x^{(\text{S})}\}$
 with respective sizes $|\x^{(k)}|$, $k\in\{1,\hdots \text{S}\}$, the parameter vector $\theta$
can be learned by minimizing
\begin{align*}
-\log \mathcal{L}(\mathcal{D};\theta) & =
-\sum_{k=1}^{\text{S}} \sum_{\t=2}^{|\x^{(k)}|} 
\log \p_\theta(x^{(k)}_{\t+1}|{\x^{(k)}_\t}).
\end{align*}
We learn the parameters using the Slashdot dataset, which consists of $\text{S} = 9,820$~threads,
containing more than $2\cdot 10^6$ comments among $93,638$ users.
In Slashdot, the most relevant feature is the preferential attachment, as detailed in~\cite{gomez13www}.
This will have implications in the optimal control solution, as we show later.

\subsection{Control interaction}
\label{controlinteraction}
The control interaction is done by highlighting a single comment of the conversation.
We assume a behavioural model for the user inspired by~\cite{Craswell}, where the user looks at the highlighted comment and decides to reply or not.
For simplicity, we assume that the user chooses the highlighted comment with a fixed probability $p'~=~\alpha/(1+\alpha)$ and with probability $1-p'$ she chooses 
to ignore it. If the highlighting of the comment is ignored, the thread grows according to the uncontrolled process.
Therefore, $\alpha$ parametrizes the strength of the influence the controller has on the user. 
For $\alpha \rightarrow \infty$, we can fully control the behaviour and for $\alpha=0$, the thread evolves according to the uncontrolled process.
A typical control would have a small $\alpha$ as usually the influence of an controlling agent on a social systems is weak.

\subsection{Experimental Setup}
To evaluate the proposed framework we use a simulated environment, without real users.
We consider a finite horizon task with $\T = 50$ with the goal to maximize the h-index at end-time,
starting from a thread with a single node as initial condition.
The state-space consists of $50!\approx 3^{64}$ states.
The thread grows in discrete time-steps. At each time-step, a new node is added to the thread by a (simulated) user.
For that, we first choose which node to highlight (optimal action) as described in section~\ref{sec:limited} using equation~\eqref{eq:ustarproxy}.
We then simulate the user as described in section \ref{controlinteraction}, so the highlighted node is selected with probability
$p'~=~\alpha/(1+\alpha)$ as the parent of the new node.
Otherwise, with probability $1-p'$, the user ignores the highlighted node and the parent of the new node is chosen according to the Slashdot model, equation~\eqref{eq:model}.
This is repeated until the end time.

\subsection{Experimental Results}
\label{sec:resultsthreads}


We first analyse the performance of the adaptive importance sampling algorithm described in section~\ref{sec:sampling}
for different fixed values of $\lambda$.

\begin{figure}[!t]
\centering
\includegraphics[width=.6\columnwidth]{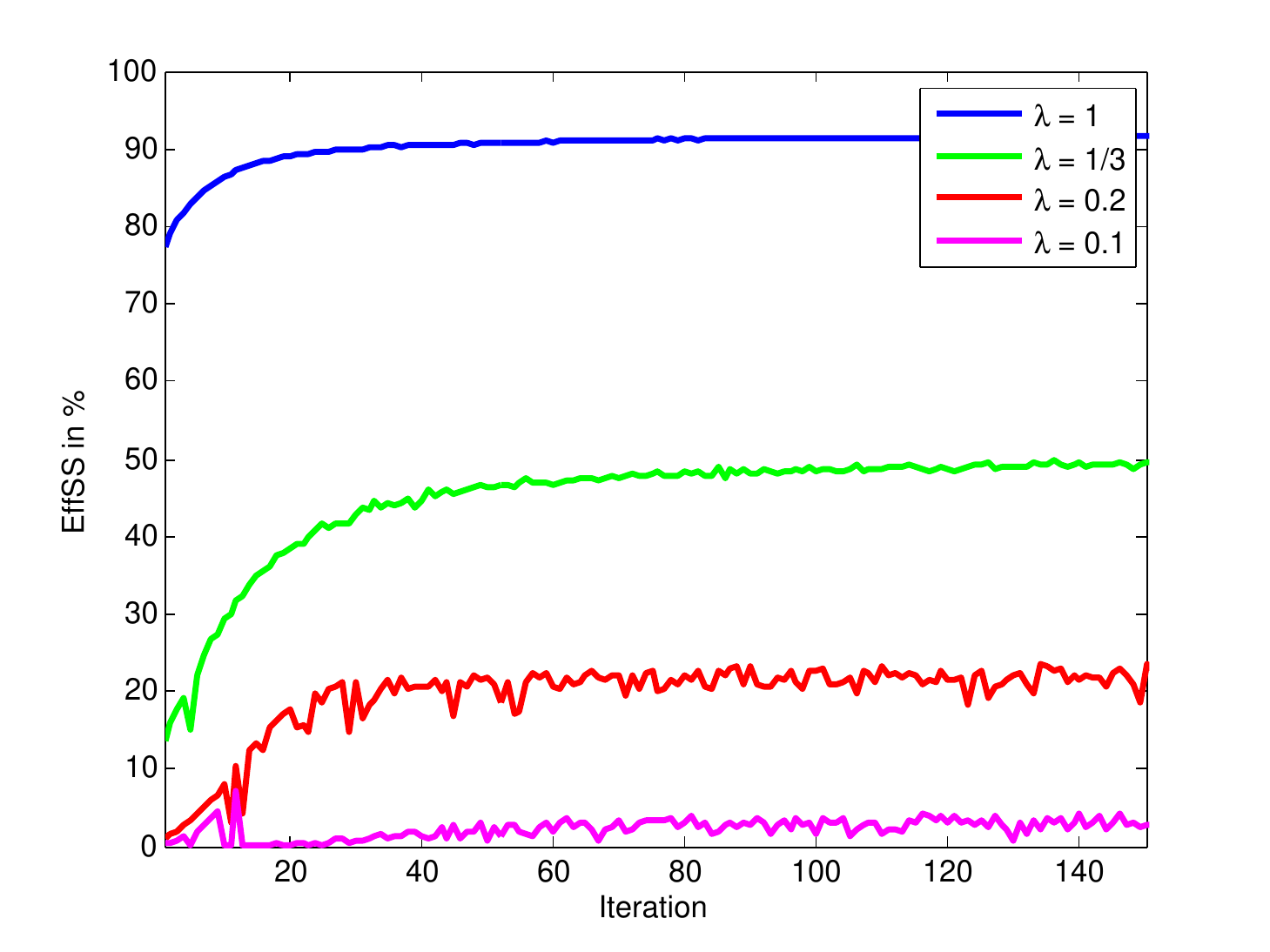}
\caption{\label{fig:EffSS} 
\textbf{Evaluation of the inference step}: The Effective sampling size (EffSS) 
increases after several iterations of the cross-entropy method.
As expected, large values of the temperature $\lambda$ result in higher values of EffSS.
We use $M=10^5$ samples to compute the EffSS. The EffSS is measured here in percent of the maximum number of samples $M$.
}
\end{figure}
 
Figure~\ref{fig:EffSS} shows the effective sample size (EffSS), equation~\eqref{eq:EffSS} as a function of the number of iterations of the CE method. We observe that the EffSS increases to reach a stable value.
As expected, large temperature (easier) problems result in higher values of EffSS.
We can also see that, even for hard problems with low temperature, the obtained EffSS is significantly larger than zero, which allows us to
compute the $\KL$-optimal control.
In general, the curves are less smooth for smaller values of $\lambda$, because a few qualitatively better samples
dominate the EffSS, resulting in higher variance.
On the other hand we also observe that the EffSS never reaches 100$\%$.
This is expected, as this would mean that our parametrized importance sampler perfectly resembles the optimal control,
and this is not possible due to the approximation error introduced by the use of features.

\begin{figure}[!t]
\centering 
\includegraphics[width=0.6\columnwidth]{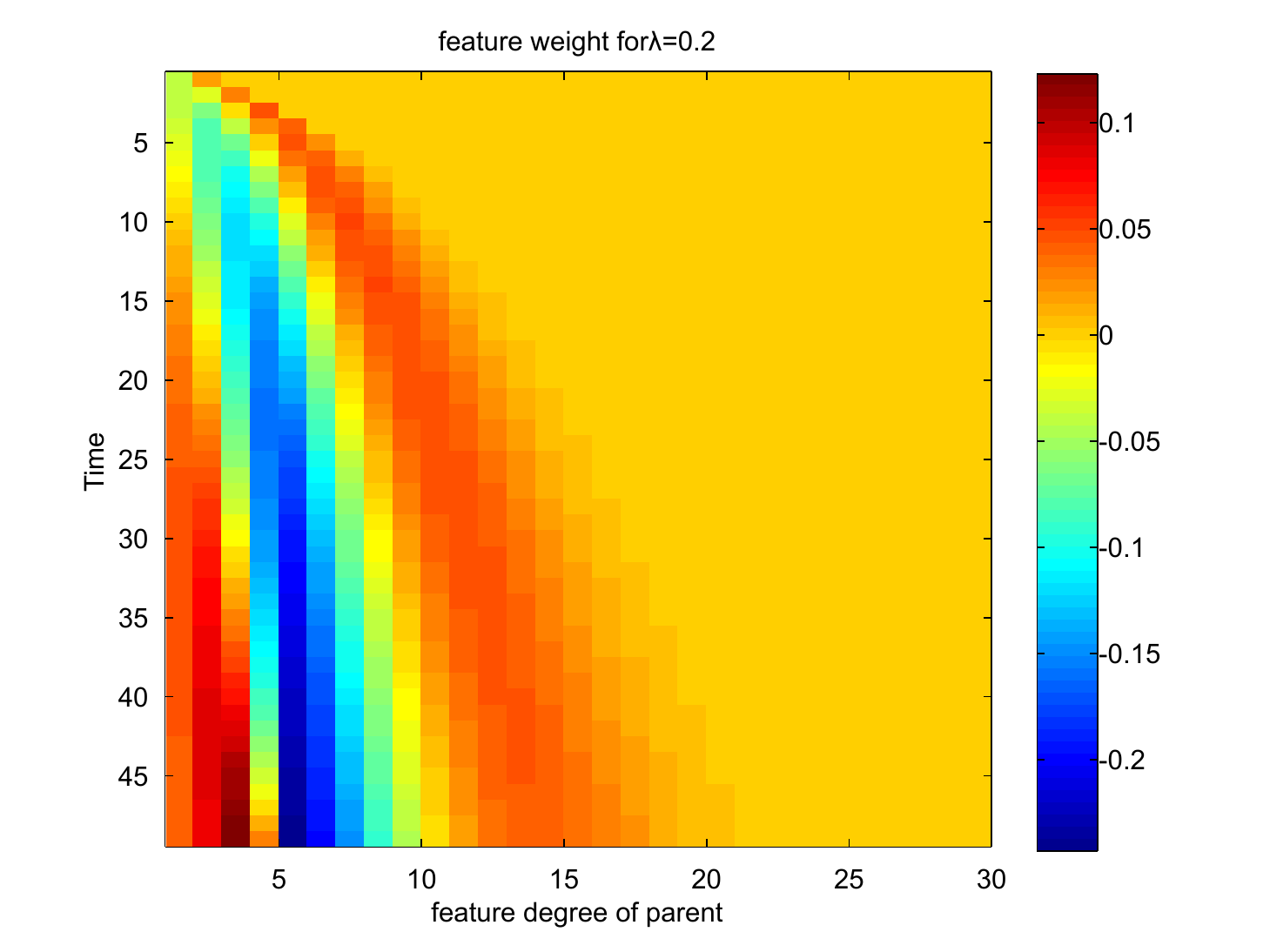}
\protect\caption{\label{fig:featuremap} \textbf{The learned importance sampler}: The
figure shows the time-dependent parameters of the learned expected
cost-to-go for $\lambda=0.2$. Each pixel is the parameter of a feature at a certain
time. The features are the degrees of the parent node after the new
child attaches. The colour represents the weight of the parameter.
Large negative weights (pixels in blue colour) stand for a low cost
and thus a desirable state, while large positive weights (red pixels)
stand for high cost and thus undesirable states. At all times there
is a desirable degree which the parent should have and higher as
well as lower degrees are inhibited. This desirable degree is small
at early times and becomes larger at later times. }
\end{figure}

We can better understand the learned control by analysing the linear 
coefficients of the parametrized optimal cost-to-go, equation~\eqref{eq:Jparam}, for this problem.
Figure~\ref{fig:featuremap} shows the feature weights $\omega_k(\t)$, at different times $\t=1,\hdots,\T$,
after convergence of the CE method.
Feature $k$ corresponds to the number of nodes with degree $k$ in the tree,
after a new node arrives. The parent node to which the new node has attached is thereby the only node whose degree changes (the degree increases by $1$). Thus a high weight for a feature which measures the number of nodes with a certain degree $k$ results in a low probability of attaching to a node with degree $k-1$. Conversely low, or large negative weights thus correspond to nodes which have a high probability of becoming the parent of the next node which is added.
We observe that there is an intermediate preferred degree (large negative weight, in blue).
This is the preferred degree of the parent of the new node, and this preferred degree increases with
time, reaching a value of $5$ at $\t = 50$.

Does this strategy make sense? 
The maximum h-index of a tree of $50$ nodes is $7$, and it is achieved if $6$ nodes have exactly $7$ children and one node has $8$. However, achieving such a configuration requires a very precise control.
For example, increasing too much the degree of a node, say up to $9$, prevents the maximum h-index to be reached,
as there are not enough links left, due to the finite horizon.
Thus, in this setting, steering for the maximal possible h-index is not optimal.
The controller prefers all parents to have a degree of $5$ and not less, but also not much more. As having more than five parents with degree at least five will result in an h-index of 5 we conclude that the control seems to aim for a target h-index of $5$, while preventing \emph{wasting} links to higher or lower degree nodes, which would not contribute to achieve that target.

The interpretation of why the preferred degree increases with time involves the uncontrolled dynamics. Remember that the most relevant term in equation~\eqref{eq:model} for the considered dataset corresponds to the preferential attachment,
parametrized by $\alpha$.
This term boosts high-degree nodes to get more links.
If this happens, most of the links end up attached to a few parents, and this effect can only be suppressed by a strong control.
The controller prevents that self-amplifying effect by aiming initially for an overall low degree,
preventing a high impact of the preferential attachment.
This keeps the process controllable and allows for a more equal distribution of the links.

\begin{figure}[!t]
\centering 
\includegraphics[width=0.65\columnwidth]{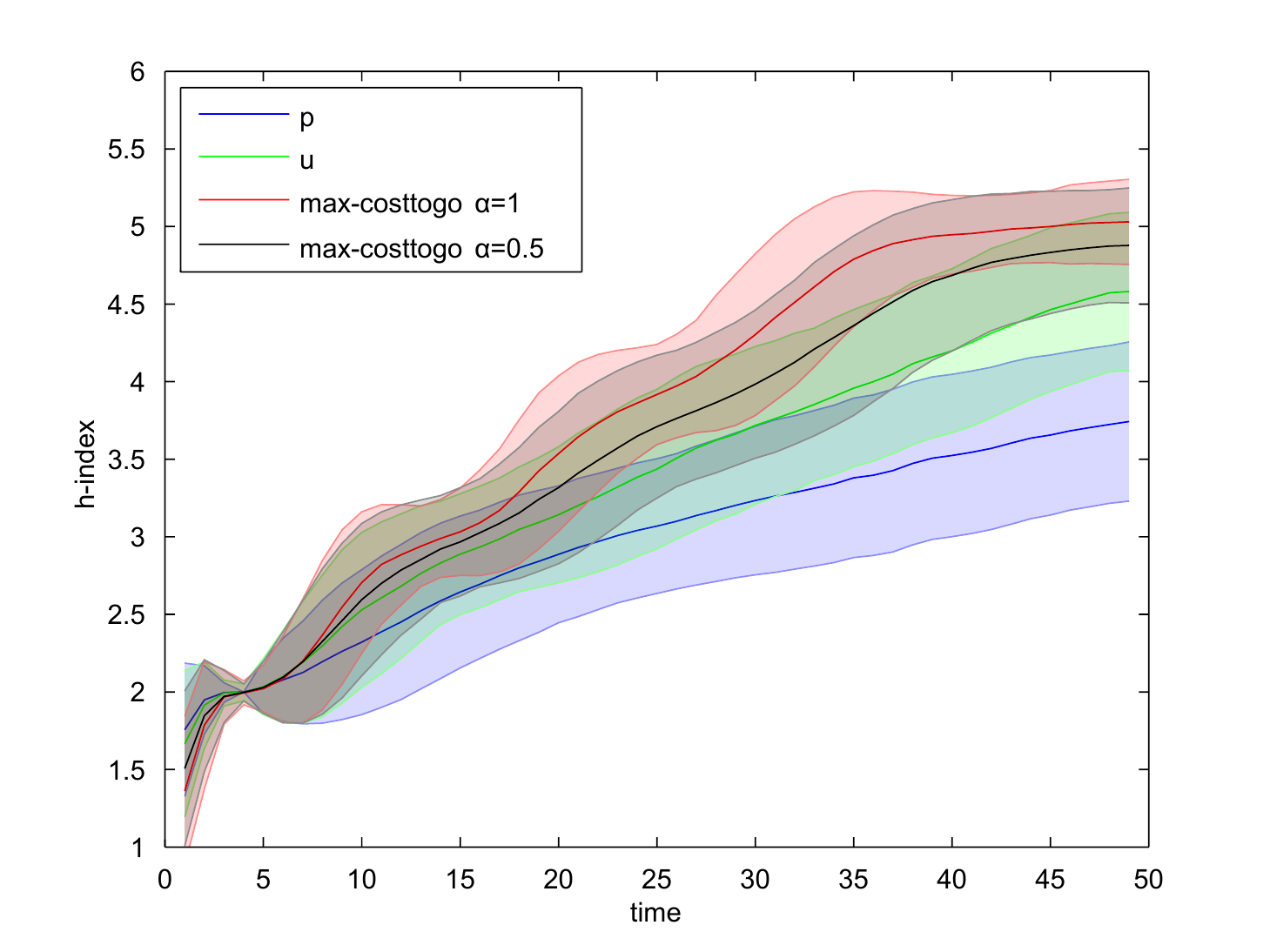}
\protect\caption{\label{fig:h-index-limited} \textbf{Evaluation of the actual control}:
uncontrolled dynamics (blue), $\KL$-optimally controlled dynamics (green)
action selection based control for $\alpha=1$ (red) and $\alpha=0.5$ (black).
The $\KL$-optimally controlled dynamics, which optimize the sum of the $\lambda$-weighted $\KL$-term and the end cost, shifts
the final mean value from about $3.7$ to about $4.7$. 
The action selection based control, which is aiming to optimize the end cost only,  is able to shift the h-index to even higher values then the $\KL$-optimal control.
For the controlled dynamics, $\lambda=0.2$ for all
three cases. 
To compute the control in each time-step we sample $1000$ trajectories.
The statistics where computed using $1000$ samples for
each of the three cases.}
\end{figure}

After having evaluated the sampling algorithm, we evaluate the proposed mechanism for actual control of the
conversation thread.
As described in section \ref{sec:limited}, in our simulated scenario, 
we highlight the node as the parent which minimizes the computed expected cost-to-go. 

Figure~\ref{fig:h-index-limited} shows the evolution of the h-index using different control mechanisms.
The blue curve shows how the h-index changes under the uncontrolled dynamics.
On average, it reaches a maximum of about $3.7$ after $50$ time steps.
In green, we show the evolution of the h-index under a $\KL$-optimal controlled case,
for temperature $\lambda=0.2$. 
As expected, we observe a faster increase, on average, than using the uncontrolled dynamics.
The maximum is about $4.7$.

The red and black curves show the evolution of the h-index using the control mechanism described in subsections~\ref{sec:limited} and~\ref{controlinteraction},
where we select actions using the expected cost-to-go $\J^\lambda_\KL$ of the $\KL$-optimal control with $\lambda=0.2$, for $\alpha=1$ and $\alpha=0.5$, respectively.
In both cases the obtained h-index is even higher than the one obtained with the $\KL$-control relaxation.
Therefore, the objective for this task, to increase the h-index, can be achieved through our action selection strategy.
As expected, a stronger interaction strength $\alpha=1$ leads to higher h-indices than a lower strength $\alpha=0.5$.

\if false
The first case is showing how the h-index changes under the uncontrolled dynamics (blue line). Staring with a low value, the h-index, averaged over $1000$ trial trajectories grows over time, to reach a maximum of about $3.7$ after $50$ time steps.
The green line represents the dynamics of the h-index under a $\KL$-optimal controlled case with a temperature $\lambda=0.2$, where we assumed that we can directly influence the transition probabilities of the dynamics. Here the increase of the h-index is faster then in the uncontrolled dynamics and reaches a maximum of about $4.7$.
Finally we look at a control interaction as described in section \ref{controlinteraction} where we selected the actions using the expected cost-to-go $\J^\lambda_\KL$ of $\KL$-optimal control with $\lambda=0.2$
We evaluated this for two levels of the interaction strength $\alpha$. In both cases the final h-index is even higher as the $\KL$-optimal control. Thus the goal of the control strategy, the increase of the h-index, can be achieved using our action selection strategy.
 As expected the final h-index achieved with the stronger interaction strength $\alpha=1$ is higher, then with the lower strength $\alpha=0.5$.
\fi

\begin{figure*}[!t]
\centering 
\includegraphics[width=0.9\textwidth]{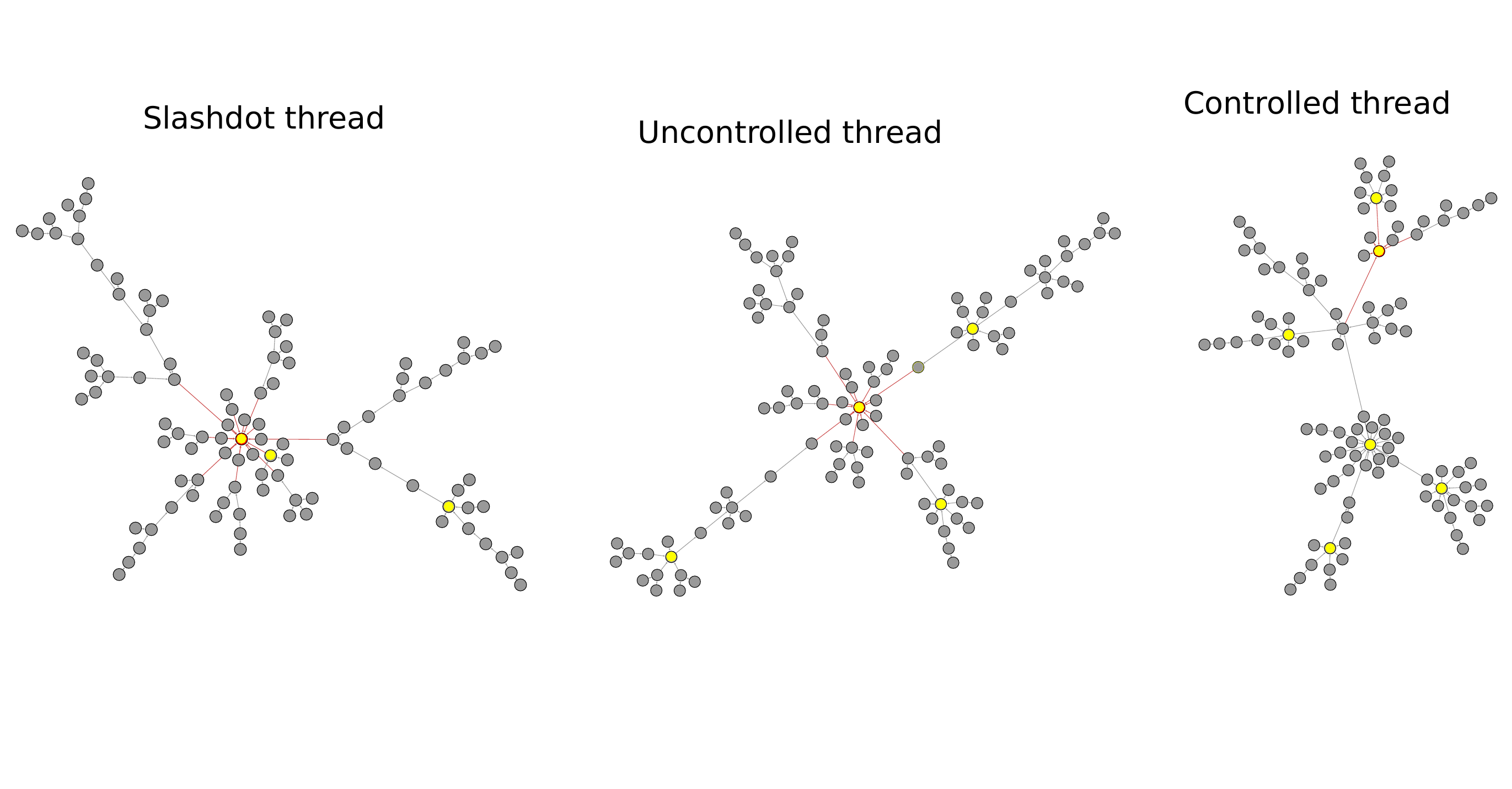}
\protect\caption{\label{fig:threads} Examples of threads. A thread from the data (Slashdot),
an uncontrolled thread generated from the model and a controlled thread.
The nodes that contribute to the h-index are coloured in yellow. The h-index for
the data and the uncontrolled thread is $4$ and $6$ for the controlled
one. }
\end{figure*}

Finally, in Figure~\ref{fig:threads} we show examples of a real discussion thread from the dataset (Slashdot), 
a thread generated from the learned model (uncontrolled process) and one resulting from applying
our action selection strategy. The latter has higher h-index.

\section{Discussion}
We have addressed the problem of controlling the growth process of a network using stochastic optimal control with the objective to optimize a structural cost that depends on the topology of the growing network. The main difficulty of such a problem is the exploding size of the state space, which grows (super-)exponentially with the number of nodes in the network and renders exact dynamic programming infeasible.

We have shown that a convenient way to address this problem is using KL-control, where a regularizer is introduced which penalizes deviations from the natural network growth process. One advantage of this approach is that the optimal control can be solved by sampling. The difficulty of the sampling is controlled by the strength of the regularization, which is parametrized by a temperature parameter $\lambda$: for high temperatures the sampling is easy, while for low temperatures, it becomes hard. This is in contrast to standard dynamic programming, whose complexity is directly determined by the number of states and independent of $\lambda$.

In order to tackle the more challenging low temperature case, we have introduced a feature-based parametrized importance sampler and used adaptive importance sampling for optimizing its parameters. This allows us to sample efficiently in the low temperature regime.
For control problems which cannot directly be formulated as $\KL$-control problems, we have proposed to use the solution of a related KL-control problem as a proxy to estimate the effective values of possible next network states. These expected effective values are subsequently used in a greedy strategy for action selection in the original control problem.
This action selection mechanism benefits from the sparsity induced by the optimal KL-control solution.


We have illustrated the effectiveness of our method on the task of influencing the growth of conversation cascades. Our control seeks to optimize the structure of the cascade, as it evolves in time, to maximize the h-index at a final time. This task is non-trivial and characterized by a sparse, delayed reward, since the h-index remains constant during most of the time, and therefore a greedy strategy is not possible.

Our approach for controlling network growth is inspired in recent approaches to optimal decision-making with information-processing constraints~\cite{TodorovPNAS2009,tishby2011information,KappenML2012,theodorou2012relative,rawlik2012stochastic}.
The Cross-Entropy method has been explored previously in the continuous case~\cite{kappen2015adaptive}.
The continuous formulation of this class of problems has been used in robotics, using parametrized policies~\cite{theodorou2010generalized,levine2013guided,gomez2014policy}.
In economics, the question of altering social network structure in order to optimize utility has been addressed mainly from a game theoretical point of view, under the name of strategic network formation~\cite{jackson2002evolution,bloch2007formation}. To the best of our knowledge, the problem of network formation has not yet been addressed from a stochastic optimal control perspective. 

The standard approach to address the problem of controlling a complex, networked system is to directly try to control the dynamics \textit{on} the network~\cite{liu2011controllability,Cornelius2013}.
This approach considers the classical notion of structural controllability as the capability of being driven from any initial state to any desired final state within finite time.
Optimal control in thus referred to the situation where a network can be fully controlled using only one driving signal.
This idea is also prevalent in the influence maximization problem in social networks~\cite{kempe2003maximizing,farajtabar2014shaping,farajtabar2015coevolve}, 
which consists in finding the subset of driver (most influential) nodes in a network.

Since the controllability of the dynamics \textit{on} the network depends crucially on the topology, several works
have considered the idea of changing the network structure is some way that favours structural controllability.

For example, the perturbation approach introduced in~\cite{Optimizing} looks for the minimum number of links that
needs to be added so that the perturbed network can be fully controlled using a single input signal.
In~\cite{hou2015enhancing}, a method to enhance structural controllability of a directed network by changing the direction of a small fraction of links is proposed.
More recently,~\cite{wang2016effective} analyzed node augmentation of directed networks while insisting that
the minimum number of drivers remains unchanged.

The main difference between our approach and these approaches is that, rather than considering the
controllability of the dynamical system \textit{on} the underlying network,
our optimal control task is defined on~\textit{the structure} of the network itself, regardless of the dynamical system
defined on it.
In some sense, our results complement these approaches. 
For example, one could use our optimal control approach to shape the growth of the network in a way that
the structural controllability, understood as the state cost function, is optimized.

\section*{Acknowledgements} 
This project is co-financed by the Marie Curie FP7-PEOPLE-2012-COFUND Action, Grant agreement no: 600387, the Marie Curie Initial Training Network ‘NETT’, project N. 289146 and the Spanish Ministry of Economy and Competitiveness under the Mar\'ia de Maeztu Units of Excellence Programme (MDM-2015-0502).

\appendix
\section{Adaptive Importance Sampling for $\KL$-Optimal Control Computation using the Cross-Entropy method}
\label{sec:tachan}
Here we show how the time-dependent weights $\omega_{k}(\t)$ of the importance sampler
are updated such that $\widetilde{\u}_{\omega}(\x'|\x,\t)$
becomes closer to the optimal sampling distribution. This corresponds to the second step of the Cross-Entropy method described in subsection~\ref{sec:sampling}.
For clarity in the derivations, we will replace $\p(\x\trajnull|\x,0)$ and $\u_\KL^{\ast}(\x\trajnull|\x,0)$ by $\avgp$ and $\avgustar$, respectively, in the expectations.
The closeness
of the two distributions $\widetilde{\u}_{\omega}(\x'|\x,\t)$ and $\u_\KL^{\ast}(\x'|\x,\t)$
can be measured as the cross entropy between the path $\x\trajnull$ probabilities under these two Markov processes:
\begin{align}
\KL\left[\u_\KL^{\ast}(\x\trajnull|\x,0)\parallel\widetilde{\u}_{\omega}(\x\trajnull|\x,0)\right]&=\left\langle \log\frac{\u_\KL^{\ast}(\x\trajnull|\x,0)}{\widetilde{\u}_{\omega}(\x\trajnull|\x,0)}\right\rangle _{\avgustar}
\notag\\
&=-\left\langle \log\widetilde{\u}_{\omega}(\x\trajnull|\x,0)\right\rangle _{\avgustar}+const.=:-\text{D}(\omega),\label{eq:CrossEntropy}
\end{align}
 where the constant term $\left\langle \log\u_\KL^{\ast}(\x\trajnull|\x,0)\right\rangle _{\avgustar}$ is dropped.

We minimize equation~\eqref{eq:CrossEntropy} by gradient descent.
At iteration $l$, the gradient $\text{D}(\omega^{(l)})$ with respect to $\omega_{k}(\t)$ is given by
\begin{align}
\frac{\partial\text{D(\ensuremath{\omega^{(l)}})}}{\partial\omega_{k}(\t)} &=-\left\langle \frac{\partial}{\partial\omega_{k}(\t)}\log\widetilde{\u}_{\omega^{(l)}}(\x\trajnull|\x,0)\right\rangle _{\avgustar}\notag
\end{align}
where 
\begin{align}
\widetilde{\u}_{\omega^{(l)}}(\x\trajnull|\x,0) &= \frac{1}{\text{Z}}\p(\x\trajnull|\x,0)\prod_{\t=0}^{\T-1}\exp\left(-\frac{\widetilde{\J}_\KL(\x_{\t+1},\omega(\t))}{\lambda}\right)
\notag \\
\text{Z} &= \left\langle \prod_{\t'=0}^{\T-1}\exp\left(-\frac{\widetilde{\J}_\KL(\x_{\t'+1},\omega(\t'))}{\lambda}\right)\right\rangle _{\avgp} \notag
\end{align}
with the normalization constant $\text{Z}$.
This leads to
\begin{align}\label{eq:gradient}
\frac{\partial\text{D(\ensuremath{\omega^{(l)}})}}{\partial\omega_{k}(\t)}
& =-\left\langle \frac{\partial}{\partial\omega_{k}(\t)}
\left(
\log \p(\x\trajnull|\x,0)
- \sum_{\t'=0}^{\T-1} \frac{\widetilde{\J}_\KL(\x_{\t'+1},\omega(\t'))}{\lambda}
- \log\text{Z}
\right)
\right\rangle _{\avgustar}
\end{align}
where we can drop the first term as it is independent of  $\omega(\t)$. The second term can be evaluated using the definition of $\widetilde{\J}_\KL$, equation \eqref{eq:Jparam}.

Further, plugging in $\text{Z}$ we get
\begin{align}
\frac{\partial\text{D(\ensuremath{\omega^{(l)}})}}{\partial\omega_{k}(\t)}
&=  \lambda^{-1}\left\langle \psi_{k}^{\t}(\x_{\t+1})\right\rangle _{\avgustar}+
\frac{\partial}{\partial\omega_{k}(\t)}
\left\langle
\log
\left\langle \prod_{\t'=0}^{\T-1}\exp\left(-\frac{\widetilde{\J}_\KL(\x_{\t'+1},\omega(\t'))}{\lambda}\right)\right\rangle _{\avgp}
\right\rangle _{\avgustar}
\notag\\
&=  \lambda^{-1}\left\langle \psi_{k}^{\t}(\x_{\t+1})\right\rangle _{\avgustar}+
\frac{\partial}{\partial\omega_{k}(\t)}
\log
\left\langle \prod_{\t'=0}^{\T-1}\exp\left(-\frac{\widetilde{\J}_\KL(\x_{\t'+1},\omega(\t'))}{\lambda}\right)\right\rangle _{\avgp}
\notag\\
&=  \lambda^{-1}\left\langle \psi_{k}^{\t}(\x_{\t+1})\right\rangle _{\avgustar}+
\frac{
1
}
{
\text{Z}
}
\frac{\partial}{\partial\omega_{k}(\t)}
\left\langle \prod_{\t'=0}^{\T-1}\exp\left(-\frac{\widetilde{\J}_\KL(\x_{\t'+1},\omega(\t'))}{\lambda}\right)\right\rangle _{\avgp}
\notag\\
&=  \lambda^{-1}
\left(
\left\langle \psi_{k}^{\t}(\x_{\t+1})\right\rangle _{\avgustar}-
\frac{1}
{
\text{Z}
}
\left\langle \psi_{k}^{\t}(\x_{\t+1})\prod_{\t'=0}^{\T-1}\exp\left(-\frac{\widetilde{\J}_\KL(\x_{\t'+1},\omega(\t'))}{\lambda}\right)\right\rangle _{\avgp}
\right)
\notag\\
&=  \lambda^{-1}
\left(
\left\langle \psi_{k}^{\t}(\x_{\t+1})\right\rangle _{\avgustar}-\left\langle \psi_{k}^{\t}(\x_{\t+1})\right\rangle _{\widetilde{\u}_{\omega^{(l)}}(\x\trajnull|\x,0)}
\right)\notag\\
& = \lambda^{-1}
\left(
\frac{\left\langle \frac{\p(\x\trajnull|\x,0)}{\widetilde{\u}_{\omega}(\x\trajnull|\x,0)}\phi\left(\x\trajnull\right)\left(\psi_{k}^{\t}(\x_{\t+1})\right)\right\rangle _{\widetilde{\u}_{\omega^{(l)}}(\x\trajnull|\x,0)}}{\left\langle \frac{\p(\x\trajnull|\x,0)}{\widetilde{\u}_{\omega}(\x\trajnull|\x,0)}\phi\left(\x\trajnull\right)\right\rangle _{\widetilde{\u}_{\omega^{(l)}}(\x\trajnull|\x,0)}}-\left\langle \psi_{k}^{\t}(\x_{\t+1})\right\rangle _{\widetilde{\u}_{\omega^{(l)}}(\x\trajnull|\x,0)}
\right),
\end{align}
where we have used the estimates from the importance sampling step and equation (\ref{eq:optcontrol}).
%

\begin{algorithm}[t]
\caption{Cross-Entropy Method for KL-control}\label{alg:alg}
\begin{algorithmic}
\REQUIRE importance sampler $\widetilde{\u}_{\omega}$,\\
 $\qquad \qquad $feature space $\psi(\cdot)$,\\
		$\qquad \qquad $number of samples $\M$,\\
		$\qquad \qquad $learning rate $\eta$\\
\STATE $l\leftarrow0$
\STATE $\omega_{k}^{(l)}(\t)\leftarrow0 \text{,  Initialize weights for all $k$, $\t$, $l$}$ 
\STATE $\x^{(i)}\traj \leftarrow $ draw $\M$ sample trajectories $\sim \widetilde{\u}_{\omega^{(l)}}$,
$i = 1,\hdots,\M$
\REPEAT
\STATE compute gradient $\frac{\partial\text{D(\ensuremath{\omega^{(l)}})}}{\partial\omega_{k}(\t)}$ using
 equation~\eqref{eq:gradient}
\STATE $\omega_{k}^{(l+1)}(\t)\leftarrow\omega_{k}^{(l)}(\t)+\eta\frac{\partial\text{D(\ensuremath{\omega^{(l)}})}}{\partial\omega_{k}(\t)}$ for all $k$, $\t$, $l$
\STATE $\x^{(i)}\traj \leftarrow $ draw $\M$ samples $\sim \widetilde{\u}_{\omega^{(l+1)}}$
\STATE $l\leftarrow l+1$
 \UNTIL{convergence}
\end{algorithmic}
\end{algorithm}
The update rule for the parameters becomes 
\begin{equation}
\omega_{k}^{(l+1)}(\t)=\omega_{k}^{(l)}(\t)+\eta\frac{\partial\text{D(\ensuremath{\omega^{(l)}})}}{\partial\omega_{k}(\t)}, \label{eq:adaptimp}
\end{equation}
for some learning rate $\eta$.
Algorithm \ref{alg:alg} summarizes the CE method applied to this context.

\section{Analyzing the KL-optimal cost-to-go based action selection}
\label{sec:analyzinglimited}
We have introduced an action selection framework which is based on an approximation of the optimal cost-to-go $\J(\x',\t)$ by the optimal cost-to-go $\J^\lambda_\KL(\x',\t+1)$ of a parametrized family of $\KL$-control problems which 
share the same state cost $\r(\x,\t)$.

Why is this a good idea? Consider the two extreme cases where the temperature $\lambda$, which parametrizes the family of equivalent $\KL$-control problems, is zero or infinite, respectively.

\paragraph{Extreme case $\lambda\rightarrow0$ (zero temperature):}
The total cost in the $\KL$-control problem becomes equal to the total cost in the original control problem, equation \eqref{eq:totalCost}, as the $\KL$ term vanishes.
The $\KL$-optimal control becomes deterministic: 
\begin{equation}
\lim_{\lambda\rightarrow0}\u^{\ast}_\KL(\x'|\x,\t)=\lim_{\lambda\rightarrow0}\frac{\p(\x'|\x)\exp\left(-\frac{\J^\lambda_\KL(\x',\t+1)}{\lambda}\right)}{Z}
=\begin{cases} 1   &\text{for $\x'=\text{argmin}~\J^\lambda_\KL(\x',\t+1)$}\\
								0 &\text{otherwise}
								\end{cases},
\end{equation}
where $Z$ is a normalization constant.

Thus, for $\lambda\rightarrow0$, the $\KL$-control problem becomes identical to the original problem
if the system is fully controllable, i.e. for every $\t$, $\x$ and $\tilde{\x}$ there is a $\u_{\tilde{\x}}\in\mathcal{U}$ such that $\p(\x'|\x,\t,\u_{\tilde{\x}})=\delta_{\tilde{\x},\x'}$.

\paragraph{Extreme case $\lambda\rightarrow\infty$ (infinite temperature):}
For this case, using equation~\eqref{eq:J_solution} we get 
\begin{align*}
\J_\KL^\infty(\x,\t)&=\lim_{\lambda\rightarrow\infty}\J_\KL^\lambda(\x,\t) \\
 &=\r(\x,\t)-\lim_{\lambda\rightarrow\infty}\lambda\log\left(\left\langle \exp\left(-\lambda^{-1}\sum_{\t'=\t+1}^{\text{T}}\r(\x_{\t'},\t')\right)\right\rangle _{\p(\x\traj|\x,\t)}\right)\\
 & =  \r(\x,\t)+\left\langle \sum_{\t'=\t+1}^{\text{T}}\r(\x_{\t'},\t')\right\rangle _{\p(\x\traj|\x,\t)}.
\end{align*}
Using equation \eqref{eq:totalCost} and the definition of the uncontrolled dynamics, we can write 
\begin{equation}
\J_\KL^\infty(\x,\t)=\r\left(\x,\t\right)+\left\langle \sum_{\t'=\t+1}^{\T}\r\left(\x_{\t'},\t'\right) \right\rangle _{P(\x\traj|\x,0,\t)}=\mathcal{C}\left(\x,\t,0\right).
\end{equation}
Thus, for $\lambda\rightarrow\infty$, the $\KL$-optimal cost-to-go becomes equal to the total cost in the original control
problem under the uncontrolled dynamics (using $\u=0$).
Having this equation \eqref{eq:ustarproxy} can be written as 
\begin{equation}
\u^{\ast}(\x,\t)\approx\text{argmin}_{\u}\left(\text{r}(\x,\t)+\left\langle \mathcal{C}\left(\x',\t+1,0\right)\right\rangle _{P(\x'|\x,\u,\t)}\right).\label{eq:ustarproxy2}
\end{equation}


In this case, the action selection is equivalent to optimize an expected total cost assuming the system will evolve according to the free dynamics in the future.
Thus the infinite temperature control can be used if one wants to guarantee that the obtained solution will not be worse than the solution obtained with zero control. 
Choosing a lower $\lambda$, however, might in practice work better (as we also have shown in section~\ref{sec:threads}) but has no theoretical guarantee.


We can conclude that our action selection strategy is meaningful in the two extreme cases, $\lambda\rightarrow\infty$ and  $\lambda\rightarrow0$.
Also this analysis suggests that, if the available set of actions $\u\in\mathcal{U}$ offers a strong control over the system dynamics, 
it is more convenient to use a $\J_\KL^\lambda$ with a low temperature $\lambda$.

\bibliographystyle{abbrvnat}
\bibliography{mybib}

\end{document}